\documentclass[conference]{IEEEtran}
\IEEEoverridecommandlockouts
\usepackage{balance}
\usepackage{pifont}
\usepackage{stmaryrd}
\usepackage{optidef}
\usepackage{amssymb}
\usepackage{mathrsfs}
\usepackage{amsfonts}  
\usepackage[bookmarks=false]{hyperref}
\usepackage[dvips]{color}
\usepackage{xcolor}
\usepackage{epsfig,amssymb,amsmath,amsthm,bm}
\usepackage{graphicx}
\usepackage{algorithm}
\usepackage{cuted}
\usepackage{float}
\usepackage{algorithm}
\usepackage{algpseudocode}
\usepackage{cite}

\usepackage{booktabs}

\usepackage{dsfont}
\usepackage{esint}
\usepackage{subfig}
\usepackage{subcaption}
\usepackage{url}
\usepackage{relsize}\usepackage{subdepth}\allowdisplaybreaks

\usepackage{cite}

\newcommand{\beq}{\begin{equation}}
\newcommand{\eeq}{\end{equation}}
\newcommand{\beqn}{\begin{eqnarray}}
\newcommand{\eeqn}{\end{eqnarray}}

\def\BibTeX{{\rm B\kern-.05em{\sc i\kern-.025em b}\kern-.08em
    T\kern-.1667em\lower.7ex\hbox{E}\kern-.125emX}}

\begin{document}

\title{Enhancing Off-Grid One-Bit DOA Estimation with Learning-Based Sparse Bayesian Approach for Non-Uniform Sparse Array
\thanks{
The work of Y.\ Hu and S.\ Sun was supported in part by U.S.\ National Science Foundation (NSF) under Grants CCF-2153386 and ECCS-2033433. 
 The work of Y.\ D.\ Zhang was supported in part by NSF under Grant ECCS-2236023.}}

\author{
	\IEEEauthorblockN{ Yunqiao~Hu$^{\dagger}$, Shunqiao~Sun$^{\dagger}$, and Yimin D.~Zhang$^{\S}$}\\
\IEEEauthorblockA{
$^{\dagger}$Department of Electrical and Computer Engineering, The University of Alabama, Tuscaloosa, AL 35487\\	
$^{\S}$Department of Electrical and Computer Engineering, Temple University, Philadelphia, PA 19122
}}

\maketitle

\vspace{0.5em}
\begin{abstract} 
This paper tackles the challenge of one-bit off-grid direction of arrival (DOA) estimation in a single snapshot scenario based on a learning-based Bayesian approach. Firstly, we formulate the off-grid DOA estimation model, utilizing the first-order off-grid approximation, incorporating one-bit data quantization. Subsequently, we address this problem using the Sparse Bayesian based framework and solve iteratively. However, traditional Sparse Bayesian methods often face challenges such as high computational complexity and the need for extensive hyperparameter tuning. To balance estimation accuracy and computational efficiency, we propose a novel Learning-based Sparse Bayesian framework, which leverages an unrolled neural network architecture. This framework autonomously learns hyperparameters through supervised learning, offering more accurate off-grid DOA estimates and improved computational efficiency compared to some state-of-the-art methods. Furthermore, the proposed approach is applicable to both uniform linear arrays and non-uniform sparse arrays. Simulation results validate the effectiveness of the proposed framework.  
\end{abstract}

\smallskip
\section{Introduction}

The problem of direction of arrival (DOA) estimation is fundamentally important in sensor array signal processing with widely application in radar, sonar, navigation, and wireless communications\cite{sedighi2021performance,sun2020mimo,4350230,7870764,9429942}. Most super-resolution DOA estimation algorithms such as MUltiple SIgnal Classification (MUSIC) \cite{schmidt1982signal} and Estimation of Signal Parameters via Rotational Invariant Techniques (ESPRIT) \cite{32276} have been primarily developed and applied to uniform linear arrays (ULAs), where sensor elements are arranged in a straight line with equal spacing, typically half the signal wavelength. However, in practical applications, achieving higher resolution with ULA requires a larger aperture and thus an increased number of array elements, significantly raising hardware costs\cite{7870764}. Furthermore, ULAs are susceptible to mutual coupling effects, which can degrade DOA estimation performance\cite{singh2013mutual}. To address this problem, sparse linear arrays (SLAs) have been used over the past few decades to achieve desired apertures with fewer active elements. Some SLA configurations, such as the minimum redundancy array
(MRA)\cite{4541350}, the nested array\cite{5456168}, the co-prime array\cite{5609222}, and the generalized coprime array configurations \cite{qin_2015}, 
have been well studied and analyzed in the past decades. While subspace-based methods, such as MUSIC and weighted subspace fitting \cite{liu_spl_2015,97999}, and covariance matrix-based compressive sensing methods \cite{qin_2015,8472789} can be applied to SLAs, they require a high number of snapshots to achieve accurate covariance matrix estimation, making them impractical in snapshot-limited scenarios commonly encountered in automotive applications \cite{9429942}. 

In antenna array systems, high-precision, high-sampling-rate analog-to-digital converters (ADCs) are costly and power-intensive\cite{9399770}. Low-bit or even one-bit quantization offers a cost-effective solution to simplify sampling hardware and increase sampling rates\cite{1039405}. Consequently, low-resolution ADC signal processing has drawn significant research interest, with research on one-bit DOA estimation becoming one of the key focuses. Subspace-based methods exploiting one-bit quantized data, along with performance analyses, are presented for both ULAs and SLAs \cite{sedighi2021performance}, \cite{yu_2016,8700277}. 

In the past decade, methods leveraging sparse reconstruction \cite{1468495} and compressed sensing (CS) \cite{1614066} have emerged to address one-bit DOA estimation, leading to various estimators such as the \textit{Binary Iterative Hard Thresholding} (BIHT) algorithm \cite{6418031}, the \textit{Bayesian Compressed Sensing} (BCS) algorithm \cite{6963346}, \textit{Sparse Learning via Iterative Minimization} (SLIM), and \textit{Iterative Adaptive Approaches} (IAA) \cite{9399770}. These techniques operate effectively with SLAs in single-snapshot scenarios. All the above methods are considered on-grid approaches, as they determine DOAs by peak searching over a fixed discrete angular spectrum. 

For off-grid source signals, grid refinement is often necessary to maintain estimation accuracy, necessitating denser grids and increased computational complexity. In CS-based methods, denser grids also lead to higher correlations among dictionary atoms, thereby degrading algorithm performance. To this end, gridless and off-grids estimation methods have been proposed to address an one-bit off-grid DOA estimation problem. Gridless methods, such as \textit{Atomic Norm Minimization} (ANM) \cite{zhou2017gridless}, estimate DOAs without grid division but require computationally intensive semidefinite programming (SDP). On the other hand, off-grid methods, such as \textit{off-grid iterative reweighted} (OGIR) algorithm \cite{10024794}, alternatively refine on-grid spectrum estimates and grid gap estimates for improved estimation accuracy. Off-grid methods generally eliminate the need for dense grid division making them more flexible and applicable for real DOA estimation tasks compared to gridless methods. However, these methods often require hundreds of iterations to achieve satisfacory results, with each iteration involving computationally matrix inversion.

Recently, model-based deep learning has gained traction in the signal processing community\cite{10056957}, including DOA estimation research \cite{10266765,10289861,10348517}. Such hybrid approaches combine the interpretability of classical array signal models with the representation power of deep neural networks, thus better addressing limitations of traditional methods in handling high-dimensional, noisy, or complex data.

In this paper, we propose a novel learning-based sparse Bayesian approach to tackle the problem of one-bit off-grid DOA estimation. We choose the sparse Bayesian framework for its robustness to noise and superior reconstruction accuracy, especially in single-snapshot scenarios. We first formulate the one-bit off-grid model using a first-order grid approximation. By applying the maximum a posteriori (MAP) criterion and incorporating sparse signal priors, we construct an iterative minimization problem, which is mapped to a neural network architecture, employing convolutional neural networks (CNNs) to replace matrix inversion operations and deep neural networks (DNNs) for off-grid updates. Simulation results demonstrate that the proposed method outperforms state-of-the-art algorithms in terms of accuracy, providing more precise off-grid DOA estimates across various signal-to-noise ratio (SNR) scenarios, using only a single snapshot of data.

\section{Problem Formulation}\label{sigm}
\subsection{Signal Model}

Consider a scenario involving $K$ narrowband, far-field source signals, denoted as $s_k(t)$ for $k = 1, \dots, K$, arriving at an $N$-element SLA from directions $\bm{\theta} = [\theta_1, \cdots, \theta_K]^{\mathrm T}$, where $(\cdot)^{\mathrm T}$ represents transpose. The array signal model with one-bit quantizated data is expressed as:
\begin{equation}
\begin{aligned}
\mathbf{y}(t) &= \mathrm{csgn}\left(\sum_{k=1}^K \mathbf{a}(\theta_k)s_k(t) + \mathbf{n}(t)\right)\\
&= \mathrm{csgn}\left(\mathbf{A}(\bm{\theta})\mathbf{s}(t) + \mathbf{n}(t)\right), \quad t = 1, \cdots, T,
\end{aligned}
\end{equation}
where $\mathbf{y}(t)$ is the received signal vector, $\mathbf{A}(\bm{\theta})=\left[\mathbf{a}(\theta_{1} ),\mathbf{a}(\theta_{2} ), \cdots, \mathbf{a}(\theta_{K} )\right]$ is the array manifold matrix, $\mathbf{s}(t)$ is the source signal vector, and $\mathbf{n}(t)$ is the complex Gaussian noise vector. The complex sign function is defined as $\mathrm{csgn}\left(\cdot\right)=\mathrm{sign} (\Re\left (\cdot\right)) + j\mathrm{sign}\left(\Im\left(\cdot\right)\right)$, where $\mathrm{sign}(\cdot)$ returns value in 
$\left \{1, -1\right\} $, $\Re(\cdot)$ and $\Im(\cdot)$ respectively return the real and imaginary parts of a complex number. Each column in array manifold matrix $\mathbf{A}(\bm{\theta})$ corresponds to a steering vector, given for the $k$-th signal as:
\begin{align}
\mathbf{a}(\theta_k) = \left[1, e^{j\frac{2\pi d_2}{\lambda}\sin{\theta_k}}, \dots , e^{j\frac{2\pi d_N}{\lambda}\sin{\theta_k}}\right]^{\mathrm T},
\end{align}
where $d_n$ specifies the spacing between the $n$-th element and the first element. This paper focuses on estimating the signal DOAs, $\bm \theta$, using a single-snapshot data vector $\mathbf{y}$. With $T$ set to $1$, the model simplifies to:
\begin{equation}
\mathbf{y} = \mathrm{csgn}\left(\mathbf{A}(\bm{\theta})\mathbf{s} + \mathbf{n}\right).\label{signle_snapshot_one_bit_data_model}
\end{equation}

\subsection{One-Bit Off-Grid Single-Measurement Vector Model}

To estimate $\bm{\theta}$ from $\left(\ref{signle_snapshot_one_bit_data_model}\right)$, we reformulate it as a single-measurement vector model:
\begin{equation}
\mathbf{y} = \mathrm{csgn}(\mathbf{\mathcal{A}}(\tilde{\bm{\theta}})\mathbf{x} + \mathbf{n}).\label{svm_model}
\end{equation}
where $\mathbf{\mathcal{A}}=[\mathbf{a}(\tilde{\theta}_{1}),\mathbf{a}(\tilde{\theta}_{2}), \dots, \mathbf{a}(\tilde{\theta}_{M})]\in \mathbb{C}^{N \times M}$ is the dictionary matrix, $\Theta = \{\tilde{\theta}_{1},\tilde{\theta}_{2}, \dots, \tilde{\theta}_{M}\}$ are the discretized angle grids and $\mathbf{x}=\left[x_{1}, x_{2},\dots, x_{M}\right]^{T}$ are sparse coefficients to be estimated. 

Under this model, $M$ grid points serve as the basis for sparse signal representation, and the signals are assumed to align with $K$ grid points, i.e., $\theta_k \in \Theta$, $k=1, \cdots, K$. This approach is known as the on-grid model. However, in practice, true DOAs often don't align perfectly with the predefined grid, i.e., $\theta_k \notin \Theta$. Assuming the grid is sufficiently dense, the true DOA $\theta_k$ lies near the fixed grid point $\tilde{\theta}_{n_{k}}$, $n_{k} \in \left\{1,2,\dots,M\right\}$. Using the first-order Taylor expansion, the true DOA can then be approximated as:
\begin{equation}
\theta_k = \tilde{\theta}_{n_{k}} + (\theta_k - \tilde{\theta}_{n_{k}}),
\end{equation}
where $(\theta_k - \tilde{\theta}_{n_{k}})$ represents the off-grid gap. The steering vector $\mathbf{a}(\theta_k)$ can be approximated as:
\begin{equation}
\mathbf{a}(\theta_k) = \mathbf{a}(\tilde{\theta}_{n_{k}}) + \mathbf{b}(\tilde{\theta}_{n_{k}})(\theta_k - \tilde{\theta}_{n_{k}}),
\end{equation}
where $\mathbf{b}(\tilde{\theta}_{n_{k}})=\frac{\partial \mathbf{a}(\theta)}{\partial \theta}|_{\tilde{\theta}_{n_{k}}}$ is the first-order derivative. By incorporating the approximation error into the measurement noise, the measurement model can be reformulated as:
\begin{equation}
\mathbf{y} = \mathrm{csgn}\left(\mathbf{C}(\bm{\beta})\mathbf{x} + \mathbf{n} \right), \label{one_bit_off_grid_single_snpashot_DOA_model}
\end{equation}
where $\mathbf{C}(\bm{\beta})=\mathbf{\mathcal{A}} + \mathbf{\mathcal{B}} \ \mathrm{diag}\left(\bm{\beta}\right)$ is the approximation dictionary with $\mathrm{diag}(\cdot)$ denoting diagonal matrix, 
$\mathcal{B} = [\mathbf{b}(\tilde{\theta}_{1}),\mathbf{b}(\tilde{\theta}_{2}), \dots, \mathbf{b}(\tilde{\theta}_{M} )]$, and $\bm{\beta} = \left[\beta_1,\beta_2, \dots, \beta_M\right]^{\mathrm T}$ are off-grid gaps defined as:
\begin{equation}
    \beta_n = \begin{cases}
    \theta_k - \theta_{n_k},\ \textrm{if} \ n=n_k, k\in \left\{ 1,2,...,K \right\},
 \\
  0,\ \hspace{3em} \textrm{otherwise}.
\end{cases}
\end{equation}

\section{Algorithm Framework}

\subsection{Sparse Bayesian Formulation}
To establish a sparse Bayesian framework for the estimation of one-bit off-grid DOA estimation, we first introduce a probabilistic model to quantify the probability of $\mathbf{x}$ given the input $\mathbf{y}$. According to (\ref{one_bit_off_grid_single_snpashot_DOA_model}), the posterior probability of $\mathbf{x}$, under the MAP criterion, is determined by the likelihood function $p\left(\mathbf{y}|\mathbf{x};\bm{\beta}\right)$ and the prior probability density function (PDF) $p\left(\mathbf{x}\right)$. The likelihood function is given by\cite{8642940}:
\begin{align}  
p\left(\mathbf{y}|\mathbf{x};\bm{\beta}\right) = \prod_{m=1}^{M} & \Phi\left(\frac{\Re\left(y_m\right)\Re(\bm{c}_m^\mathrm{T} \left(\bm{\beta}\right)\mathbf{x})}{\sigma/\sqrt{2}}\right) \nonumber \\
 & \cdot \Phi\left(\frac{\Im\left(y_m\right )\Im(\bm{c}_m^\mathrm{T} \left(\bm{\beta}\right)\mathbf{x})}{\sigma/\sqrt{2}}\right),
\label{likelihood function1}
\end{align}
where $\Phi\left(\cdot\right)$ denotes the cumulative density function of the standard normal distribution, $y_m$ is the $m$-th element in $\mathbf{y}$, $\bm{c}_m^\mathrm{T}$ is the $m$-th row vector in $\mathbf{C}$.  For the convenience of calculation, let $\hat{\mathbf{x}}=\frac{\sqrt{2}}{\sigma}\mathbf{x}$, and (\ref{likelihood function1}) can be reformulated as: 
\begin{align}  
p\left(\mathbf{y}|\hat{\mathbf{x}};\bm{\beta}\right) = \!\prod_{m=1}^{M} & \Phi(\Re(y_m)\Re(\bm{c}_m^\mathrm{T} (\bm{\beta})\hat{\mathbf{x}})) 
\Phi(\Im(y_m)\Im(\bm{c}_m^\mathrm{T} (\bm{\beta})\hat{\mathbf{x}})).
\label{likelihood function2}
\end{align}
A suitable prior PDF for $\hat{\mathbf{x}}$, such as a Laplacian prior \cite{5256324} or exponential distribution \cite{9399770}, should be selected to encourage sparsity. In this paper, we select the prior PDF as: 
\begin{equation}
p\left(\hat{\mathbf{x}}\right) = \prod_{i=1}^{M}\mathrm{exp}\left(-\frac{\lambda \left|\hat{x}_i\right|^{\alpha}}{\alpha}\right), \ \ 0<\alpha\le 1,
\label{piorx}
\end{equation}
where $\lambda$ is a parameter. As $\alpha$ approaches $0$, $p\left(\hat{\mathbf{x}}\right)$ reaches its maximum at $\hat{\mathbf{x}}=0$, enforcing sparsity on the signal. 
The off-grid gaps $\bm{\beta}$ follow a uniform distribution, $p\left(\bm{\beta}\right) \sim U\left(-\frac{r}{2}, \frac{r}{2}\right)$\cite{6320676}, where $r$ denotes the grid interval size. With Bayes rule, the MAP estimator is given as: 
\begin{align}
\left\{\hat{\mathbf{x}}^{*}, \bm{\beta}^{*}\right\} = \mathrm{arg}\min_{x, \beta}-\mathrm{ln}p\left(\mathbf{y}|\hat{\mathbf{x}};\bm{\beta}\right) - \mathrm{ln}p\left(\hat{\mathbf{x}}\right) - \mathrm{ln}p\left(\bm{\beta}\right).
\label{MAP estimator}
\end{align}
\vspace{-0.5cm}
\subsection{Updating Formula}
By substituting (\ref{likelihood function2}) and (\ref{piorx}) into equation (\ref{MAP estimator}), we obtain the following cost function to be minimized:
\begin{align}
\mathcal{L} = \sum_{m=1}^{M} & \left\{ -\mathrm{ln}\Phi\left(\Re\left(y_m\right )\Re(\bm{c}_m^\mathrm{T}(\bm{\beta})\hat{\mathbf{x}}) \right) \right. \nonumber\\ 
& \ \ -\mathrm{ln}\Phi \left(\Im\left(y_m\right)\Im(\bm{c}_m^\mathrm{T}(\bm{\beta})\hat{\mathbf{x}}) \right) \nonumber \\ & \ \  \left.+\sum_{i=1}^{N}\frac{\lambda\left|x_i\right|^{\alpha}}{\alpha} + \textrm{const}\right\}.
\label{obj_func}
\end{align}

Since the object function in (\ref{obj_func}) is non-convex, we apply convex relaxation to simplify it. Specifically, with the majorization-minimization (MM) principle, we can find the upper bound of the first two terms in (\ref{obj_func}) as \cite{8642940}:
\begin{align}
&\sum_{m=1}^{M}\!\left\{\!-\mathrm{ln}\Phi(\Re\left(y_m\right )\!\Re(\bm{c}_m^\mathrm{T}(\bm{\beta})\hat{\mathbf{x}}))\!-\!\mathrm{ln}\Phi (\Im\left(y_m\right )\!\Im(\bm{c}_m^\mathrm{T}(\bm{\beta})\hat{\mathbf{x}}))\!\right\} \nonumber \\
&\le \! \sum_{m=1}^{M}\!\frac{1}{2}(\Re\left(y_m\right) \Re(\bm{c}_m^\mathrm{T} (\bm{\beta})\hat{\mathbf{x}}))^{2} \!+ \frac{1}{2}(\Im(y_m)\Im(\bm{c}_m^\mathrm{T}(\bm{\beta})\hat{\mathbf{x}}))^{2} \nonumber \\
&-\Re(v_{m}^{t})\Re(y_m)\Re(\bm{c}_m^\mathrm{T}(\bm{\beta})\hat{\mathbf{x}}) \!-\! \Im(v_{m}^{t})\Im(y_m)\Im(\bm{c}_m^\mathrm{T}(\bm{\beta})\hat{\mathbf{x}})+ c'.
\label{MM_approx}
\end{align}
where $v_{m}^{t}=\left[\Re(y_m)\Re(\tilde{v}_{m}^{t})\right] + j\left[\Im(y_m)\Im(\tilde{v}_{m}^{t}) \right]$, $\tilde{v}_{m}^{t}=d_{m}^{t}-\mathrm{I}'(d_{m}^{t})$, $d_{m}^{t}=\Re(y_m)\Re(\bm{c}_m^\mathrm{T}(\bm{\beta}^{t})\hat{\mathbf{x}}^{t})+j\Im(y_m)\Im(\bm{c}_m^\mathrm{T}(\bm{\beta}^{t})\hat{\mathbf{x}}^{t})$, and $c'$ is a constant. Here, 
superscript $(\cdot)^{t}$ indicates variables from the $t$th iteration. In addition, function $\mathrm{I}'(x)$ is defined as 
\begin{equation}
\mathrm{I}'(x)=-\frac{\mathrm{exp}\left(-{\Re{(x)}^{2}}/{2}\right)}{\sqrt{2\pi}\Phi(\Re{(x)})}-j\frac{\mathrm{exp}\left(-\Im{(x)}^{2}/2\right)}{\sqrt{2\pi}\Phi(\Im{(x)})}.
\end{equation} 

The third term in (\ref{obj_func}) can be smoothly approximated as:
\begin{align}
\sum_{i=1}^{N}\frac{\lambda\left|x_i\right|^{\alpha}}{\alpha}\approx\frac{\lambda}{\alpha}\sum_{i=1}^{N}\left(\left|x_i\right|^{2}+\eta \right)^{\frac{\alpha}{2}}.\label{smooth_approx}
\end{align}
where $\eta>0$ is a small constant, typically set to $10^{-6}$. By substituting (\ref{MM_approx}) and (\ref{smooth_approx}) into (\ref{obj_func}), we obtain a new minimization problem defined as: 
\begin{align}
\left\{\hat{\mathbf{x}}^{*}, \bm{\beta}^{*}\right\} = \mathrm{arg}\min_{\hat{x}, \beta} & \ \frac{1}{2}\left\|\mathbf{C}(\bm{\beta})\hat{\mathbf{x}} - \mathbf{v}^{t}\right\|_{2}^{2} \nonumber \\ & + \frac{\lambda}{\alpha}\sum_{i=1}^{N}\left(\left|x_i\right|^{2}+\eta\right)^{\frac{\alpha}{2}} + c',
\end{align}
where $\mathbf{v}^{t} = [v_{1}^{t}, \cdots, v_{M}^{t}]^{\mathrm T}$.
Then, we use iterative update strategy to estimate $\mathbf{x}$ and $\bm{\beta}$ as:
\begin{align}
\hat{\mathbf{x}}^{t+1} & = \left[\mathbf{C}^{\mathrm H}(\bm{\beta}^{t})\mathbf{C}(\bm{\beta}^{t}) + \lambda\bm{\Lambda}(\hat{\mathbf{x}}^{t})\right]^{-1}\mathbf{C}^{\mathrm H}(\bm{\beta}^{t})\mathbf{v}^{t},  \label{step1}
\\
\bm{\beta}^{t+1} & = 
\ \Re\left((\mathcal{B}^{\mathrm{H}}\mathcal{B})^{*}\hat{\mathbf{x}}^{t+1}(\hat{\mathbf{x}}^{t+1})^{\mathrm H}\right)^{-1} \nonumber\\
&\quad \cdot\Re\left(\mathrm{diag}\left((\hat{\mathbf{x}}^{t})^{*}\right)\mathcal{B}^{\mathrm H}\left(\mathbf{v}^{t} - \mathcal{A}\hat{\mathbf{x}}^{t}\right)\right),\label{step3}
\end{align}
where
\begin{align}
\bm{\Lambda} \left(\hat{\mathbf{x}}^{t} \right)=\mathrm{diag}\left( \left(\left|\hat{x}_1\right|^{2}+\eta \right)^{\frac{\alpha}{2}-1},\cdots, \left(\left|\hat{x}_N\right|^{2}+\eta \right)^{\frac{\alpha}{2}-1}\right). \nonumber
\end{align}

\section{Neural Network derived from the Algorithm}
In this section, we follow the algorithm unrolling paradigm to design the network architecture by mapping the iteration steps (\ref{step1})--(\ref{step3}) to customized layers. 

\subsection{Network Architecture}
\begin{figure}
\centering
\includegraphics[width=0.48\textwidth]{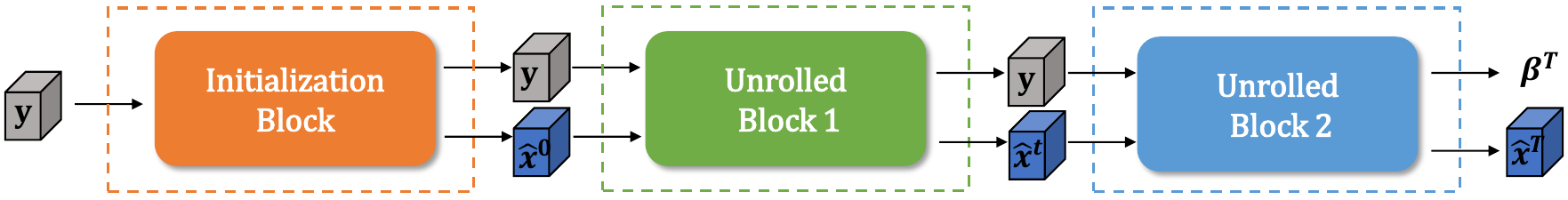}
\caption{Overall architecture of the Unrolled Network.\label{unrolled_net_overview}}
\end{figure}

The network consists of the following three main blocks: Initialization Block, Unrolled Block 1, and Unrolled Block 2, as shown in Figure \ref{unrolled_net_overview}.
\subsubsection{Initialization Block}
The initialization block simply performs the following operation on the input vector:
\begin{align}
\hat{\mathbf{x}}^{0} = \mathbf{C}^{\mathrm{H}}(\bm{\beta}^{0})\mathbf{y},
\end{align}
where $\bm{\beta}^{0}=\bm{0}$, which means that no initial off-grid gaps are assumed.
\begin{figure}
\centering
\includegraphics[width=0.38\textwidth]{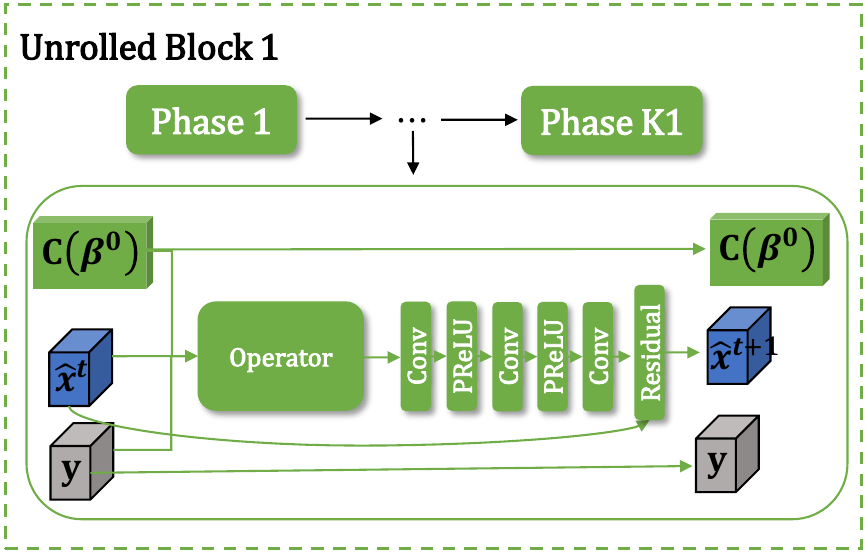}
\caption{Detail architecture of Unrolled Block 1.\label{unrolled_block1}}
\vspace{-0.5em}
\end{figure}
\subsubsection{Unrolled Block 1}
Unrolled Block 1 consists of $K_1$ unrolled phases. The details of one unrolled phase are illustrated in Figure \ref{unrolled_block1}.
Unrolled Block 1 is designed to refine the spectrum in fixed grids. Therefore, the grid gaps are set to $\bm{\beta}^{0} = \bm{0}$, and grid update is not considered in this block. The operation is defined as:
\begin{equation}
\tilde{\mathbf{v}}^{t} = \mathbf{C}^{\mathrm{H}}(\bm{\beta}^{0})\mathbf{v}^{t}, \label{initialize_layer}
\end{equation}
where $\bm{\beta}^{0}=\mathbf{0}$, $\mathbf{v}^{t}=\left[\Re\left(\mathbf{y}\right)\odot\Re(\tilde{\mathbf{v}}^{t})\right]+j\left[\Im(\mathbf{y})\odot\Im(\tilde{\mathbf{v}}^{t})\right]$, $\tilde{\mathbf{v}}^{t} = \mathbf{D}^{t} - \mathrm{I}'(\mathbf{D}^{t})$, and $\mathbf{D}^{t} = \Re\left(\mathbf{y})\odot\Re(\mathbf{C}(\bm{\beta}^{0})\hat{\mathbf{x}}^{t}\right) + j\Im(\mathbf{y})\odot\Im\left(\mathbf{C}(\bm{\beta}^{0})\hat{\mathbf{x}}^{t}\right)$, $\odot$ is Hadamard product. 
According to (\ref{step1}), matrix inversion is used to compute the estimate. However, the computational complexity of this operation increases with matrix size, making network training challenging. To address this, we replace matrix inversion in (\ref{step1}) with convolutional layers. In addition, each convolutional layer is followed by a Parametric Rectified Linear Unit (PReLU) activation layer \cite{He_2015_ICCV} to introduce nonlinearity into the network. A residual connection is incorporated into each unrolled phase to give the output $\hat{\mathbf{x}}^{t+1}$.
\subsubsection{Unrolled Block 2}
\begin{figure}[t]
\centering
\includegraphics[width=0.48\textwidth]{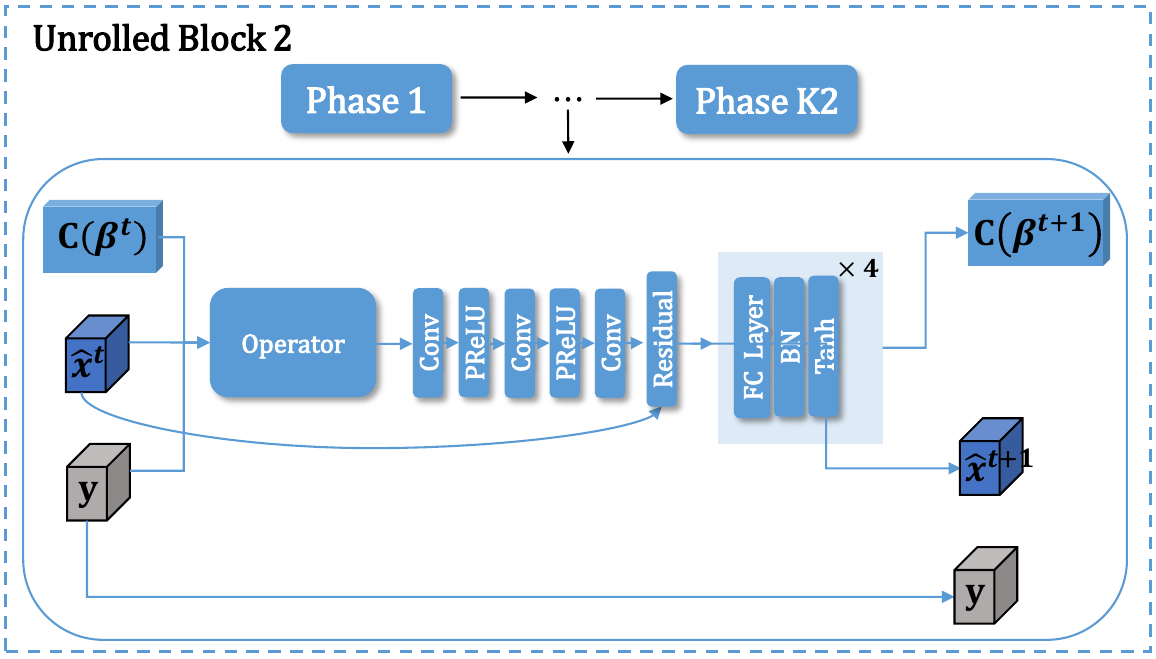}
\caption{Detail architecture of Unrolled Block 2.\label{unrolled_block2}}
\vspace{-0.5em}
\end{figure}

Unrolled Block 2 comprises $K_2$ unrolled phases and is designed to update both the angle spectrum coefficients and the off-grid gaps, as shown in Figure \ref{unrolled_block2}. Each phase follows the same operations as in (\ref{initialize_layer}), with convolutional layers identical to those in Unrolled Block 1. For off-grid gap updates, we replace the formula in (\ref{step3}) with four Fully Connected (FC) layers, each followed by batch normalization and Tanh activation. The FC layers take the absolute values of the estimated signal spectrum, $|\hat{\mathbf{x}}^{t+1}|$, as input and output the updated off-grid gaps corresponding to the on-grid spectra. The output is normalized to a single grid interval $[-\frac{r}{2}, \frac{r}{2}]$.

\begin{figure*}[htp]
\centering
\includegraphics[width=0.98\textwidth]{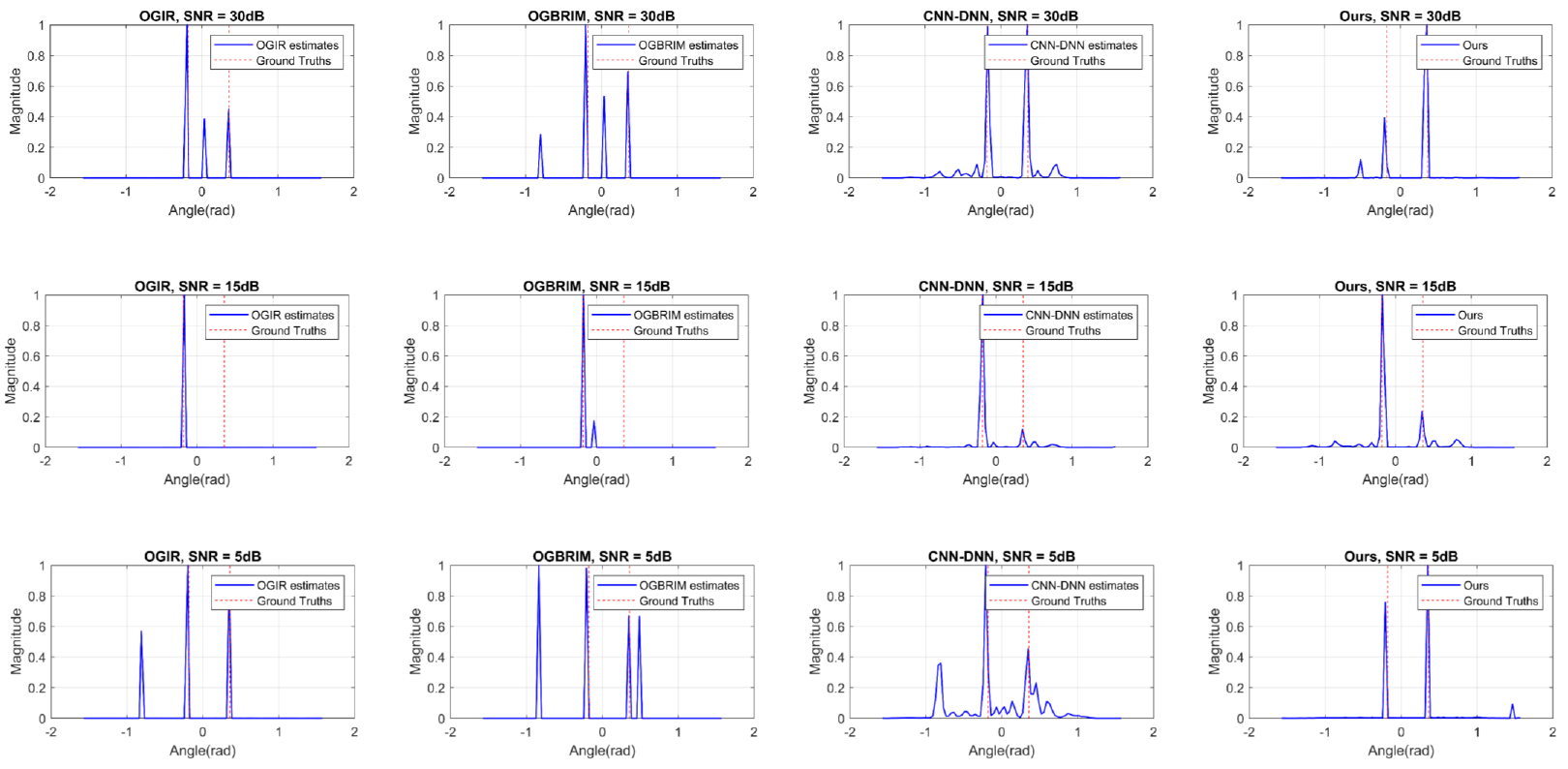}
\caption{Spectral outputs for $10$-element SLA.\label{example2}}
\end{figure*}

\subsection{Generating Datasets}\label{generate_data}
We generate training datasets using two SLA configurations with half-wavelength element spacing: an 18-element array at $\frac{\lambda}{2}[0, 1, 2, 3, 4, 7, 8, 9, 10, 11, 12, 13, 14, 15, 16, 17, 18, 19]$ and a 10-element array at $\frac{\lambda}{2}[0, 3, 4, 5, 6, 7, 11, 16, 18, 19]$. The number of signal sources is set to 2, and the array field of view (FOV) is defined as $[-60^{\circ}, 60^{\circ}]$, discretized with a fixed grid size of $2^{\circ}$. The off-grid gaps corresponding to target sources are randomly generated following a uniform distribution $U\left(-1^{\circ} , 1^{\circ}\right)$. Reflection coefficients for DOA sources are generated as random complex numbers, with their real and imaginary parts uniformly distributed in $U\left(0.5 , 1\right)$. Denoting the ground truth of the $n$-th DOA as $G_n$, the signals are labeled according to
\begin{equation}
\mathbf{s}_{n}^{*}  =
\begin{cases}
|s_k|, & \text{if } \theta_k = G_n, \\
0, & \text{otherwise.}
\end{cases}
\end{equation}
For the labeling of off-grid gaps:
\begin{equation}
\beta_{n}^{*} = \begin{cases}
|\beta_k|, & \text{if } \theta_k = G_n, \\
0, & \text{otherwise.}
\end{cases}
\end{equation}
We randomly generate $100,000$ samples across input SNR levels ranging between $0$ dB and $30$ dB in $5$ dB increments. $90\%$ of the dataset is used for training and the remaining $10\%$ is used for validation.

\subsection{Training Approach}
Our training process is divided into two stages. In the first stage, we train only Unrolled Block 1 while keeping the parameters of Unrolled Block 2 frozen. This training process is performed with a batch size of $64$ for $100$ epochs. The objective of this stage is to generate angle spectra with low sidelobe levels, facilitating the subsequent training of Unrolled Block 2. The loss function used in this stage is the Binary Cross-Entropy (BCE) loss, defined as:
\begin{equation}
    \mathcal{L}_1(\hat{\mathbf{x}}, \mathbf{s}^{*}) = - \frac{1}{N}\sum_{i=1}^{N} \left[ s_{i}^{*}\cdot\mathrm{log}\hat{x}_i + (1-s_{i}^{*})\cdot\mathrm{log}(1-\hat{x}_i) \right].
\end{equation}
In the second stage, the parameters of Unrolled Block 1 are frozen, and only Unrolled Block 2 is trained. In this stage, we apply a combination of the mean squared error (MSE) loss and the BCE loss. The total loss function is defined as:
\begin{align}
    \mathcal{L}_2(\hat{\mathbf{x}}, \mathbf{s}^{*};\hat{\bm{\beta}},\bm{\beta}^{*}) &= -\frac{1}{N}\sum_{i=1}^{N} \left[s_{i}^{*}\!\cdot\!\mathrm{log}\hat{x}_i + (1-s_{i}^{*})\!\cdot\!\mathrm{log}(1-\hat{x}_i) \right]\nonumber \\
    &\ \ \ + \frac{1}{N}\sum_{i=1}^{N}[\hat{\beta}_i - \beta_{i}^{*}]^{2}.
\end{align}

\section{Performance Evaluation}\label{perf}

In this section, we evaluate the performance of the proposed method in terms of detection rate and root mean square error (RMSE). The angular space between $-90^\circ$ and $90^\circ$ is discretized into $91$ fixed grids, each with the grid size of $2^\circ$. An angle estimation is considered successful if the absolute errors of all estimated DOAs are within a threshold of $0.5^\circ$.  Otherwise, it is considered a failure. The detection rate is defined as $\frac{N_s}{N_t}$, where $N_t$ is the total number of Monte Carlo tests and $N_s$ is the number of successful tests. The RMSE is calculated as $\mathrm{RMSE}=\sqrt{\frac{1}{N_s\hat{K}}\sum_{t=1}^{N_s}||\hat{\bm{\theta}}_t - \bm{\theta}^{*}||_{2}^{2}}$, where $\hat{\bm{\theta}}_t$ represents the estimated DOA vector in the $t$th test round. A total of $1,024$ Monte Carlo trials are conducted for testing. The evaluation considers two off-grid target sources, $\bm{\theta}^{*} = [-10.28^\circ, 20.56^\circ]$, tested with the two SLA configurations described in Section \ref{generate_data}. For comparison, we include the OGIR algorithm \cite{10024794}, representing algorithm-based methods, and the CNN-DNN network \cite{chung2021off}, representing network-based approaches. Additionally, the algorithm framework proposed in this paper, referred to as ``OGBRIM", is used as a baseline for comparison. 

\begin{figure}[h]
\centering
\includegraphics[width=0.33\textwidth]{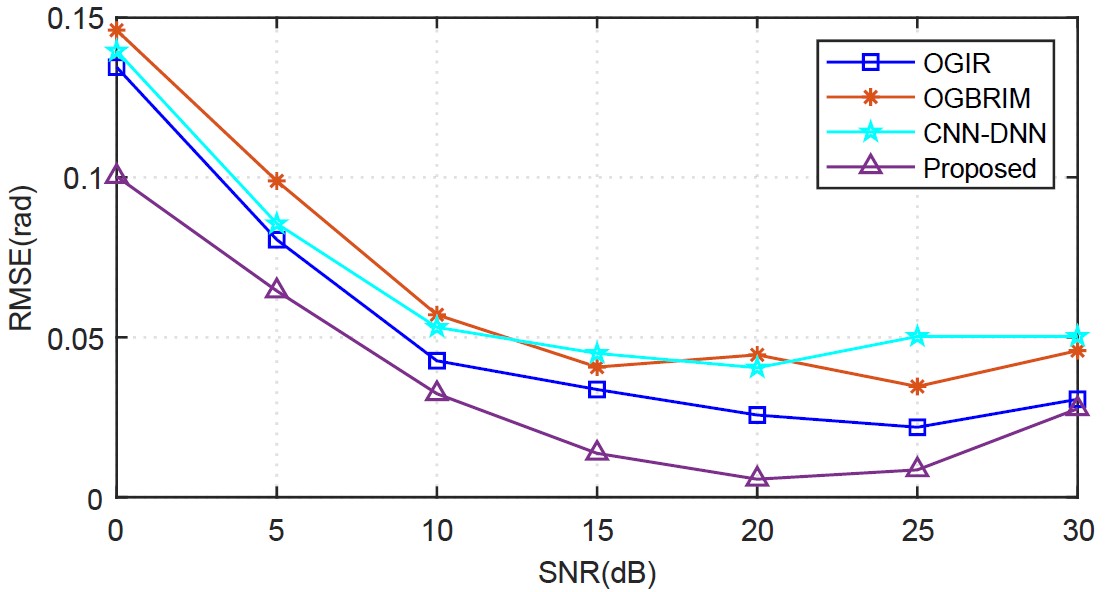}\\
(a) RMSE vs.\ inpput SNR\\
\medskip
\includegraphics[width=0.33\textwidth]{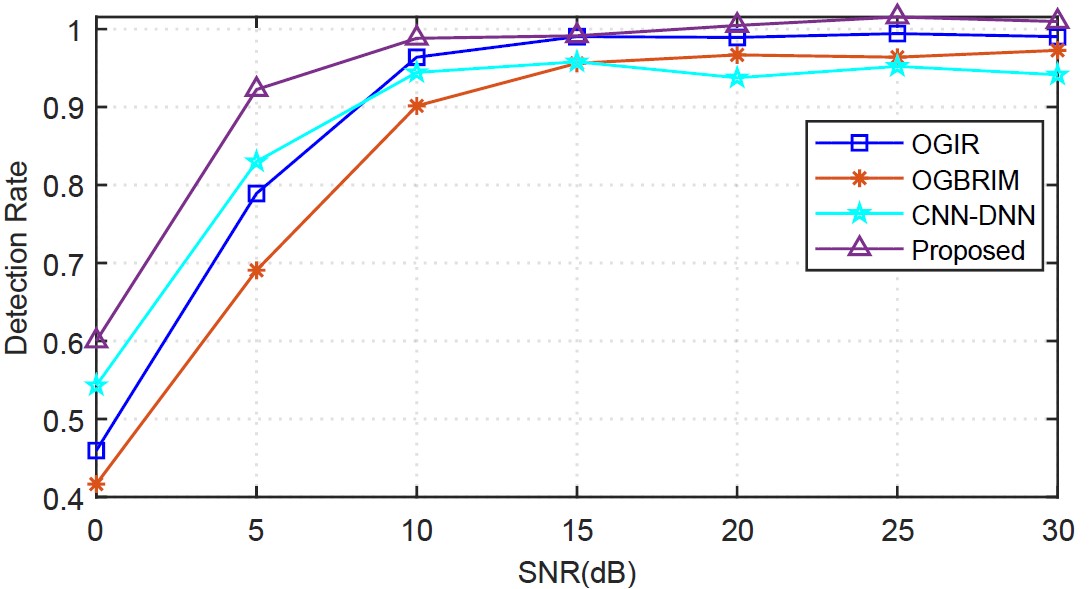}\\
(b) Detection rate vs.\ input SNR\\
\caption{$18$-element SLA. \label{RMSE1andDetect1}}
\end{figure}
\begin{figure}
\centering
\includegraphics[width=0.33\textwidth]{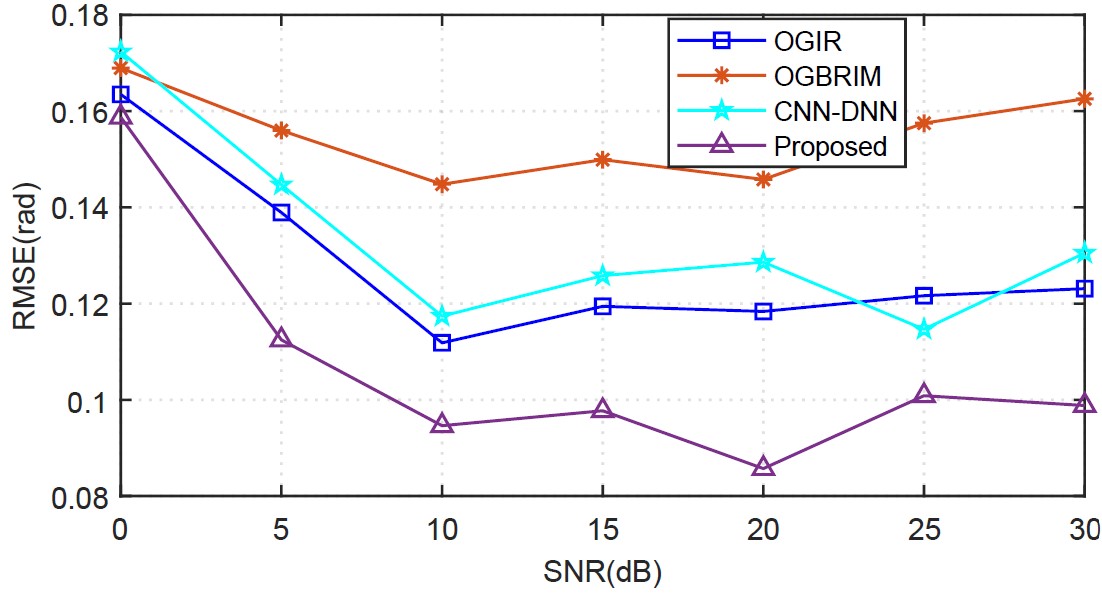}\\
(a) RMSE vs.\ input SNR\\
\medskip
\includegraphics[width=0.33\textwidth]{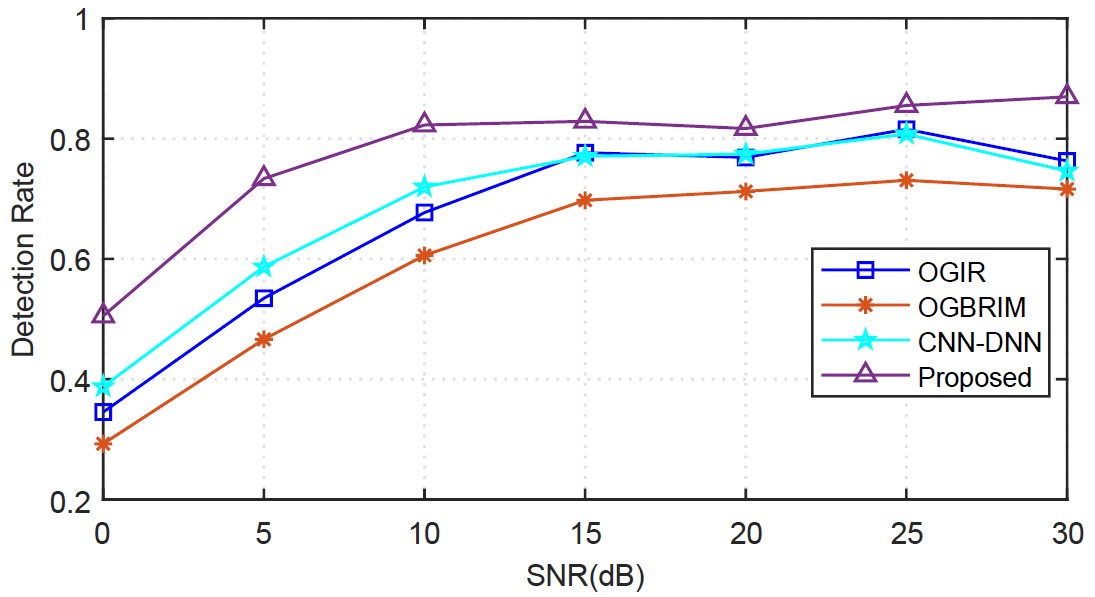}\\
(b) Detection rate vs.\ input SNR\\
\caption{$10$-element SLA. \label{RMSE2andDetect2}}
\vspace{-0.5em}
\end{figure}

We first compare the RMSE and detection rate for DOA estimation using an $18$-element SLA and a $10$-element SLA. As illustrated in Figure \ref{RMSE1andDetect1} for the $18$-element SLA, the proposed method achieves the lowest RMSE and the highest detection rate, showcasing superior estimation performance. In Figure \ref{RMSE2andDetect2}, while all methods exhibit performance degradation with the sparser $10$-element SLA, the proposed method consistently maintains lower RMSE and higher detection rates, highlighting its robustness to array sparsity.

Additionally, we compare the spectra outputs of different methods for the $10$-element SLA, and the results are depicted in Figure \ref{example2}. It is observed that the proposed method not only suppresses spurs effectively but also resolves signals across various input SNR levels. In contrast, other algorithms, such as OGIR and OGBRIM, tend to produce false estimates or miss targets.

\section{Conclusion}
In this paper, we proposed a novel learning-based sparse Bayesian approach for one-bit off-grid DOA estimation. The method effectively combines the advantages of traditional Bayesian modeling with modern neural network architectures to achieve robust and accurate off-grid angle estimation. Simulation results demonstrated that the proposed approach outperforms state-of-the-art methods across a range of SNR scenarios. Future work will explore the generalization capabilities of the proposed method to enhance its reliability and extend the approach to handle real-world radar datasets.

\bibliographystyle{IEEEtran}
 \balance
\bibliography{references}

\end{document}